\documentstyle[11pt,subeqn]{article}   
\begin{document}
\hbadness=10000

\pagestyle{plain}
\baselineskip 18.pt

\begin{center}
\centerline{\bf SINGLE PARTICLES AND COMPOSITE SYSTEMS}

\centerline{\bf IN A MATHEMATICALLY RIGOROUS FORMULATION OF}

\centerline{\bf RELATIVISTIC LAGRANGIAN QUANTUM FIELD THEORY}

               Michael Danos\footnote{Visiting Scholar}\\
Enrico Fermi Institute, University ot Chicago\\
Chicago IL, 60637\\
\end{center}

\begin{abstract}

We define quantum field theory by taking the Lagrangian action to be given as
a sequence of mathematically well-defined functionals written in terms of
operator fields fulfilling given \hbox{local} commutation relations.  The
renormalized solution fields have a fully defined Fock space expansion and are
\hbox{multi-local}; thus Haag's theorem does not apply, i.e., the interaction
picture exists.  Also, the formalism allows
immediately the definition of a wave function and the description of many-body
bound-state systems.
\end{abstract}

\bigskip

\section {INTRODUCTION}
\setcounter{equation}{0}
\renewcommand{\theequation}{1.\arabic{equation}}

In the present paper we will show that it is
possible to formulate relativistic quantum field theory (QFT) in a
mathematically rigorous way
in that every step in the chain  of reasoning, beginning with the
Lagrangian and terminating with the \hbox{S-matrix} can be
rigorously defined and that all the quantities participating in
this chain exist as fully specified mathematical objects.
In this formulation the interacting theory joins smoothly to the
non-interacting case.  At the same time the treatment of bound
composite system presents no
difficulties; in particular, it allows the definition of a system
wave function which is fully analogous to the wave function of
non-relativistic quantum mechanics.   In detail, the modifications
from the conventional
formulation of QFT are:   (i)~the basic fields have to be defined as
local quantum fields;  (ii)~the Lagrangian functional must be replaced by a
generalized functional (short: \hbox{g-functional}), which is the
analogue of the replacement of functions by generalized functions
(short: \hbox{g-functions}), see below
and Appendix~A.

Of the above mentioned two modifications, only the first is new;
the second has been implied but only rarely used explicitly.  Conventionally
it is postulated that the ``renormalized" fields are local fields.  In the
present formulation the basic fields of the Lagrangian, i.e., the so-called
``free" fields, are defined to be local fields, and to obey once and for all
fixed commutation relations. These fields are used to construct a Fock space,
and all quantities of the theory are expanded in that basis.  This way the
``renormalized" fields, i.e., the solution fields, as the result of
the interactions
acquire structure, i.e., they are non-local, and are not unitarily
related to the
``free" fields.  Hence Haag's theorem \cite{4} does not apply in our
formulation, i.e., the interaction picture exists.  This way it is
this first
of the modifications which is the basis for the resolution of the
contradictions inherent in the conventional formulation.  One may
say that the
present \hbox{re-formulation} entails the reversal of the roles of
the ``bare"
and the ``dressed" fields.

On the other hand, something along the lines of the second of the
modifications, the introduction of the \hbox{g-functional} to achieve
convergent expressions for the Feynman graphs, has been {\it implicitly}
utilized
previously. The reason for doing the calculation {\it explicitly} in terms
of a
\hbox{g-functional} rather than to regularize the divergences is to
maintain
control over the treatment so as to keep intact Noether's theorem
and the other
quantities and interrelations of the Lagrange-Hamilton based quantum theory.

It will be demonstrated that in the present formulation the Lagrangian
\hbox{g-functional} can be chosen such that without any regularization:

(a)~the Fock space is well-defined;

(b)~the propagators are of the Pauli-Villars form and are the Green's
functions
for the Lagrangian \hbox{g-functional} constructed on the Fock
space ;

(c)~the solution state vectors are given by a normalizable expansion on the
Fock space basis;

(d)~the renormalization constants, as all the other quantities of the theory,
are given unambiguously by absolutely convergent Feynman --  Schwinger
integrals;

(e)~the solution fields, i.e., the fields which in the Heisenberg picture
fulfill the equations of motion including the interactions, are
\hbox{non-local};

(f)~Haag's theorem \cite{4,5,6} does not apply \cite{7};

(g)~in the limit $\lambda\to 0$ (see below, Eq.~(\ref{1.1}) and the discussion)
the renormalized \hbox{S-matrix} coincides with that computed in the
conventional way;

(h)~the theory obeys all Ward-Takahashi identities derivable from Noether's
theorem \cite{8,9,1,2,3};

(i)~ the concept of a wave function remains valid for relativistic QFT, and
allows the description of many-body bound states;

(j)~the solutions are analytic functions of the coupling constant, say $g^2$,
as it tends towards zero;

(k)~except for locality the theory fulfills the axioms of constructive field
theory \cite{10,11}, including that the physical quantities arising in QFT are
at
most tempered \hbox{g-functions}; and

(l)~K\"allen's theorem \cite{12} applies.

As is well known, the mathematical problems of the conventional formulation of
QFT are the following:

(A)  The starting point are the solutions of the $\lambda = 0$ form (in
contrast to the $\lambda\to0$ limit) of the \hbox{g-functional} of our
formulation.  Most likely such solutions do not exist for \hbox{non-trivial}
theories.  Symptomatic for this are, for example, the statements:  the
\hbox{S-matrix} has an essential singularity at $g^2=0$; Fock space does not
exist for non-trivial QFT.

(B)   In the regularization the Green's functions get mutilated by some
prescription, be it the \hbox{Pauli-Villars} regularization or the dimensional
regularization.  Neither the regularized Green's functions, nor the remaining
(finite) parts used in the calculation are the Green's functions of the
problem at hand; they are disconnected from the assumed Lagrangian.

(C)   The Green's functions of the theory break the boundaries of tempered
\hbox{g-functions}; as has been demonstrated earlier \cite{1,2}  they contain
\hbox{meta-g-functions} (see Appendix~A).

(D)  According to the postulated properties of the (renormalized) fields, in
particular their local character, Haag's theorem does apply and renders the
results of QFT meaningless.

(E)  The problem of the convergence of the perturbation expansion of the
\hbox{S-matrix} for a generic (3+1) QFT, the empirically observed at least
semi-convergence of QED notwithstanding, remains unsolved and is not addressed
in the present paper.

In view in particular of the points (c) and (k) the mathematics
of the present formulation is very much more simple than that of the
conventional formulation.

We now give a short overview of the points which will be demonstrated
in the paper.

The generalized (operator) functional, the \hbox{g-functional}, is defined by
a sequence of functionals (see Appendix~A):
\begin{equation}
L \{\psi_i,\ldots ; \lambda_n \} ~=~ L_0 \{ \psi_i, \ldots ;
\lambda_n \} ~+~ L_1 \{ \psi_i,\ldots ; \lambda_n \}
\quad , \label{1.1}
\end{equation}
where $\lambda_n$ is a parameter (e.g., having the dimension of a
length).  The functional is constructed such that for $n\to\infty$, i.e., for
$\lambda_n\to0$, $L \{ \lambda_n \}$ approaches the  desired Lagrangian
action (say,
the quantum electrodynamics (QED) Lagrangian action), and  that for
$\lambda\neq 0$
all Feynman integrals of the Schwinger --- Tomonaga expansion of the
\hbox{S-matrix} converge.  The ``basic fields" $\psi_i$ etc., are quantized
fields, i.e., operator fields,  defined to obey given local
commutation relations.  They can be chosen to fulfill the equations of motion
resulting from $L_0 \{\cdot\}$.  The (trivial) question of normalizability of
the continuum can be taken care of in any way one likes, e.g., by the Weyl
eigendifferential method, or by periodic boundary conditions.  The basic
fields can be used to construct a Fock space.  No difficulty is encountered
in the expansion of any quantities of the full theory in this Fock basis
since, in view of the convergence of all graphs, all
such quantities are mathematically fully defined directly in the Fock space
without any mutilation by cut-offs or regularization of any integrals.
The  solutions, e.g., the expansion of the state vectors, can be obtained
either by perturbative methods, or \hbox{non-perturbatively}, by matrix
diagonalization.  This way no difficulties arise in the description of
bound states of composite systems, like, for example, positronium.

In the present paper we provide a complete chain of the steps needed for the
demonstration (of the mathematical aspects) of our formulation.  However,
we shall not give any proofs in terms
of the rigorous mathematical language; our proofs, if at all given, will be in
the simple language of a physicist.  In fact, since the mathematical concepts
arising in our formulation are rather simple, i.e., are at most tempered
distributions (tempered \hbox{g-functions}), the need for the mathematician's
language anyway is not that great.

For concreteness we conduct the discussion in terms of the simplest possible
non-trivial cases, i.e., a \hbox{g-functional} which has as its  limiting form
a spinor field interacting with a pseudoscalar field (Section~2), and the case
of spinor electrodynamics (Section~3).  The respective \hbox{g-functionals}
are written in terms of Heisenberg-picture quantized fields.  They are
manifestly relativistically covariant in general, i.e., not only in the limit
$\lambda_n\to 0$.  To continue the development we need to derive the
Hamiltonian, i.e., we must apply the variational \hbox{calculus} which in our
case we need for quantized rather than for classical fields.  As shown
in Ref.~\cite{VAR} the functional derivative, and hence the variational
calculus, can be rigorously  defined also for quantized fields, i.e., for
operator fields and that then it yields the same expressions as for
\hbox{c-number} fields.  In view of this result, the conventional procedure
which consists in splitting the Lagrangian into $L_0^{(n)}$ and $L_1^{(n)}$,
the ``free" and the ``interactions" part; in deriving the corresponding
Hamiltonian; and in going over to the interaction picture, is mathematically
fully defined and can be applied.  At this point one can, as done
conventionally, compute the \hbox{S-matrix}.   All these steps are performed
using only the basic fields. {\it The solution fields, which are supposed to
fulfill the  complete equations of
motion, are not required for computing the \hbox{S-matrix}}.

At this point one may terminate the development, as all
perturbative predictions of the theory can now be computed.  However, the
physics and the need in particular for our modification~(i) has remained
hidden. This need becomes evident in the description of the solutions,
in particular of bound many-body systems, as, for example, positronium.
Thus, to demonstrate that the modification (i) actually
is basic to the viability of QFT, we continue the
development, and we derive in
Section~4 the form of the field operators, say $\Psi$, and of the state
vectors, say $\vert W \rangle$, of a solution.  In the \hbox{g-functional}
formulation they are mathematically \hbox{well-defined} quantities.  In
particular, the solution fields can be expanded in terms of products of the
basic fields; this demonstrates the existence of the Fock space for the
interacting theory.  Furthermore, this shows that the solution fields and the
basic fields here are not unitarily equivalent; because of this Haag's theorem
does not apply.  The state vectors for any arbitrary physical system are
normalizable; so are its wave functions, defined as usually as the matrix
element $\langle 0\vert ~\Psi~\vert W\rangle$; this demonstrates the existence
of the wave function in relativistic QFT.  As in the case of the Pauli-Villars
regularization procedure, for low energy processes the auxiliary (ghost, i.e.,
negative metric and/or ``wrong" statistics, and normal) particle degrees of
freedom for $\lambda_n\to 0$ freeze out, except for providing for the
convergence of all Feynman integrals.  Owing to this convergence K\"allen's
theorem \cite{12} applies, which means that the $\lambda ~\to~0$ results
are independent of
the specific choice of the g-functional.

As for previous work, as already mentioned above, the ingredient (ii) of the
present approach, i.e., the use of modified Lagrangians, both polynomial and
non-polynomial, to achieve convergent \hbox{Feynman} integrals is very
old \cite{13,14,15,16}.  It has been employed in many papers; for example, in
more recent work devoted to the proof of the existence of the  $\phi^4$
theory \cite{17,18} the development is carried out using regularized
propagators
without actually specifying the underlying   Lagrangian; only the existence of
such a Lagrangian is postulated.

A discussion of the formulation of QFT in
terms of path integrals lies outside of the frame of this paper.

\section {YUKAWA QUANTUM FIELD THEORY}
\setcounter{equation}{0}
\renewcommand{\theequation}{2.\arabic{equation}}

The simplest possible non-trivial quantum field theory is provided by
interacting massive spin 0 and spin ${1\over2}$ fields.  This system has no
further symmetries beyond those imposed by Lorentz invariance.  The presence
of further invariances, e.g., gauge invariance, requires imposing constraints
on the solutions.  We shall treat such a case in the next Section.  Here we
consider  the minimal theory, which, however, involves the full mathematical
apparatus needed for the present \hbox{re-formulation} of QFT.

Consider the Lagrangian action \hbox{g-functional}
\begin{equation}
L \{\psi \ldots \} ~=~ L_F(\lambda) + L_B(\lambda) + L_I(\lambda)
\quad , \label{2.1}
\end{equation}
(for simplicity we write $\lambda$ instead of $\lambda_n$; also the
limit $\lambda \to 0$  is understood throughout)
\begin{eqnarray}
L_F(\lambda) &=& - \int ~ d^4x ~:~ \{ \xi_1 \,
\bar\psi_1(x) ~(\gamma\partial + m_1) ~ \psi_1(x) ~+~  \xi_2 \,
\bar\psi_2(x) ~(\gamma\partial + m_2) ~ \psi_2(x) \cr
&~& \cr
&~& \qquad \qquad +~ \xi_3 \, \bar\psi_3(x) ~(\gamma\partial + m_3) ~
\psi_3(x)\}~: \label{2.2} \\
&~& \cr
L_B(\lambda) &=& -\int ~ d^4x ~:~ \left\{ {\textstyle {1\over2}}~
\left(\partial_\mu(x)\right)^2 ~+~ {\textstyle  {1\over2}}~ \left( M\varphi(x)
\right)^2 \right\}: \label{2.3} \\
&~& \cr
L_I &=& ig \int~d^4x ~\sum_{i,j} ~ \kappa_i\kappa_j ~:~ \bar\psi_i(x) ~
\gamma_5 ~\psi_j(x) ~ \varphi(x): \quad . \label{2.4}
\end{eqnarray}
This form is manifestly relativistically invariant. It contains 2 ``auxiliary"
Fermion fields.  As we will
see, this is a possible choice; there exist infinitely many other
possible choices.

The fields $\psi_i(x),~ \varphi(x)$ are defined to obey the
commutation relations ($|\xi_i |~=~|\epsilon|~=~1$, see below)
\begin{eqnarray}
 \left[ \psi_{i\alpha}({\bf x},t), ~\psi_{j\beta}^\dagger({\bf y},t) ~+
~\epsilon ~\psi_{j\beta}^\dagger({\bf y},t) ~  \psi_{i\alpha}({\bf x},t)
\right] &=& \xi_i \, \delta_{ij} \, \delta_{\alpha\beta} \,
\delta^3({\bf x-y}) \label{2.5} \\
~&~&\cr
\left[ \varphi({\bf x},t), ~\dot\varphi({\bf y},t) \right]_- &=& i \,
\delta^3({\bf x-y}) \quad , \label{2.6}
\end{eqnarray}
where $\alpha,\beta$ are spinor indices.  Fields which have
$\xi = -1$ are ghost fields while normal fields have $\xi = +1$, and $\epsilon
= -1$ yields the ``wrong" statistics.  Further, we define
\begin{subequations}\label{2.7}
\begin{eqnarray}
m_1 &=& m \label{2.7a} \\
m_2 &=& {c_2\over\lambda} \label{2.7b}\\
m_3 &=& {c_3\over\lambda} \label{2.7c}
\end{eqnarray}
\end{subequations}
with $c_2$ and $c_3$ real positive constants.  This defines the $\lambda\to0$
limit of the \hbox{g-functional}.  The \hbox{c-number} constants $\kappa_i$
will be determined below.

Since, as shown in \cite{VAR}, variational calculus is valid not only for
\hbox{c-number} but also for operator fields one can go through the usual
procedure, i.e., use the non-interacting part of $L$ to obtain $H_0$, go over
to the interaction picture, and compute the \hbox{S-matrix} in the
standard manner (see Appendix~B).  To that end one needs the graph expansion
of the Tomonaga -- Schwinger equation.  In view of the form of (\ref{2.1}),
(\ref{2.2}), this expansion will turn out to be exactly of the Pauli-Villars
form as we now demonstrate.

Consider the Fermion Green's function as it
arises, for example, in the Wick expansion of the second order term of the
Neumann series of the Tomonaga -- Schwinger equation:
\begin{equation}
G_F ~=~ T~\sum_{jk} \, \kappa_j\kappa_k ~\langle 0|~ \psi_j(x)~ \bar\psi_k(y)~
|0\rangle \quad . \label{2.8}
\end{equation}
In view of the commutation relations (\ref{2.5}) we have in momentum space
(for brevity we omit the infinitesimal imaginary parts in the denominators)
\begin{eqnarray}
\widetilde G_F &=& K_1 ~{{\gamma p+m_1}\over {p^2 -m_1^2}} ~+~ K_2 ~{{\gamma
p+m_2}\over {p^2 -m_2^2}} ~+~ K_3 ~{{\gamma p+m_3}\over {p^2 -m_3^2}} \cr
~&~&\cr
&=& \biggl[ \gamma p~ \Bigl\{ (K_1+K_2+K_3) ~p^4 ~-~
\left[ K_1(m_2^2+m_3^2)~+~ K_2(m_1^2+m_3^2)~+~ K_3(m_1^2+m_2^2)\right] ~p^2
\Bigr.\Bigr. \cr
~&~&\cr
~&~& \qquad \bigl. +~\left[K_1m_2^2m_3^2~+~ K_2m_1^2m_3^2~+
~K_3m_1^2m_2^2\right] \Bigr\} \cr
~&~&\cr
~&~& \qquad +~\Bigl\{ (K_1m_1+K_2m_2+K_3m_3)~p^4 \bigr. \cr
~&~&\cr
~&~& \qquad +~\left[K_1m_1(m_2^2+m_3^2)~+~ K_2m_2(m_1^2+m_3^2)~+~
K_3m_3(m_1^2+m_2^2)\right]~ p^2 \cr
~&~&\cr
~&~& \qquad \Bigl.\Bigl. +~\left[K_1m_1m_2^2m_3^2~+~K_2m_2m_1^2m_3^2~+~
K_3m_3m_1^2m_2^2\right] \Bigr\} \biggr] \cr
~&~&\cr
\qquad &\times & \left[(p^2-m_1^2)(p^2-m_2^2)(p^2-m_3^2)\right]^{-1} \quad
. \label{2.9}
\end{eqnarray}
Here
\begin{equation}
K_i ~=~\xi_i ~\kappa_i^2 \quad . \label{2.10}
\end{equation}
In order to cancel the terms of the numerator containing the factors $(\gamma
p)\,p^4$   and $p^4$ there must hold
\begin{eqnarray}
\sum ~K_i &=& 0  \label{2.11} \\
\sum ~m_i\, K_i  &=& 0 \quad . \label{2.12}
\end{eqnarray}
Hence (we put $K_1~=~1$ so as to achieve the needed form for $\lambda\to0$)
\begin{subequations}\label{2.13}
\begin{eqnarray}
K_2 &=& ~-~{{c_3-\lambda m_1}\over{c_3-c_2}} \label{2.13a} \\
~&~&\cr
K_3 &=& {{c_2-\lambda m_1}\over {c_3-c_2}} \quad . \label{2.13b}
\end{eqnarray}
\end{subequations}
With these relations (\ref{2.9}) can be simplified to the expression
\begin{eqnarray}
\widetilde G_F &=&  \biggl[ \gamma p~ \left\{ \left[ K_1~ m_1^2 ~+~ K_2
~m_2^2~+~ K_3~ m_3^2 \right] ~p^2 ~+~\left[ {K_1\over{m_1^2}} ~+~
{K_2\over{m_2^2}} ~+~ {K_3\over{m_3^2}} \right] m_1^2\,
m_2^2\, m_3^2 \right\} \biggr. \cr
~&~&\cr
&~& \quad \biggl. +~\left\{ -~\left[ K_1m_1^3 ~+~ K_2m_2^3 ~+~ K_3m_3^3
\right]~ p^2   ~+~\left[ {K_1 \over{m_1}} ~+~ {K_2\over{m_2}} ~+~ {K_3 \over
{m_3}}    \right] m_1^2\, m_2^2\, m_3^2 \right\} \biggr] \cr
~&~&\cr
\qquad &\times~& \left[(p^2-m_1^2)(p^2-m_2^2)(p^2-m_3^2)\right]^{-1} \quad
. \label{2.9'}
\end{eqnarray}
Here $K_1$ has been retained so as to exhibit the symmetries of the
expression.

This way we have asymptotically $\widetilde G_F \to |p|^{-3}$. Together with
the asymptotic character $|p|^{-2}$  of the Boson propagator one sees that the
elementary Feynman graphs for the Fermion self-energy, the Boson self-energy,
and the vertex correction, have   asymptotic behavior as $|p|^{-5}$,
$|p|^{-6}$, and $|p|^{-8}$, respectively.  The integrals $\int d^4p(\cdot)$
thus are absolutely convergent, and no auxiliary (ghost) particles are
required for the Boson field.  The choice (\ref{2.2}), (\ref{2.3}) for the
\hbox{g-functional} thus turns out to be satisfactory. If
one has chosen $c_2~<~c_3$ then $\psi_2$ is a ghost field and $\xi_2 = -1$,
$\xi_1=\xi_3=1$ in (\ref{2.5}).

The fields $\psi_i(x)$ have the usual momentum space expansion:
\begin{equation}
\psi_{i\alpha}(x) ~=~ \int {{d^3p}\over {(2\pi)^{3/2}}} ~{\sqrt{m_i \over
\omega_i}} ~ \left[ b_i({\bf p}) ~u_i^{(\alpha)}({\bf p})
~e^{i({\bf px}-\omega_it)} ~+~ d_i^\dagger({\bf p}) ~v_i^{(\alpha)}({\bf p})
~e^{-i({\bf px}-\omega_it)} \right] \label{2.14}
\end{equation}
where $\omega_i = +{\sqrt{{\bf p}^2 + m_i^2}}$, and the operators $b,d$ obey
the usual, $\xi$- and $\epsilon$-modified anti-commutation relations
\begin{subequations}\label{2.15}
\begin{eqnarray}
\left[ b_i^{(\alpha)}({\bf p}), ~b_j^{(\beta)\dagger}({\bf q})
\right]_{\epsilon} &=& \xi_i \, \delta_{ij} \, \delta_{\alpha\beta} \,
\delta^3({\bf p-q})  \label{2.15a} \\
~&~&\cr
\left[ d_i^{(\alpha)}({\bf p}), ~d_j^{(\beta)\dagger}({\bf q})
\right]_{\epsilon} &=& \xi_i \, \delta_{ij} \, \delta_{\alpha\beta} \,
\delta^3({\bf p-q}) \quad , \label{2.15b}
\end{eqnarray}
\end{subequations}
and where all other anti-commutators vanish.

One now can compute the renormalization constants, i.e., the mass shifts for
the Fermion and the Boson, the vacuum-polarization shielding of the coupling
constant, and the normalization constants, which all are finite, and   proceed
to compute the \hbox{S-matrix}.  In view of the convergence of the
renormalization constants here K\"allen's theorem \cite{12} applies; thus
according to that theorem the renormalized results are independent of the
renormalization scheme, which means that all choices of
the \hbox{g-functionals} having the same limiting form and yielding convergent
Feynman graphs will give identical
results.

We conclude this section by indicating the manner in which contact
singularities arise in the Feynman -- Schwinger integrals.  To that end,
consider the
Fermion self-energy which is in second order
\begin{equation}
\Sigma ~=~ \hbox{\O} \int d^4x \, d^4y ~\sum_{ijkl} ~ \kappa_i \,
\kappa_j \, \kappa_k \, \kappa_l ~\bar\psi_i(x) \, \gamma_5 \, \Delta(x-y) \,
S_{kl}(x-y) \, \gamma_5 \, \psi_j(y)  \quad . \label{2.16}
\end{equation}
Here $\hbox{\O}$ indicates that the integration has to be over an open domain
so as not to include the contact terms \cite{1,2}, and $\bar\psi_i(x)$ and
$\psi_j(y)$ serve the role of test functions. The Wightman functions which
retain the contact singularities are computed over the full domain and
hence then $\hbox{\O}$ is to be omitted.

To evaluate this expression with the aim of localizing the most singular terms
we recall
\begin{subequations}\label{2.17}
\begin{eqnarray}
\Delta(x) &=& {i\over {4\pi^2}} ~ \left\{ {1\over {x^2- i\varepsilon}} -
{\mu^2\over4} ~ \log \left( - ~{\textstyle {1\over4}} ~ \mu^2 \, x^2 \,
\widetilde\gamma^2 + i\varepsilon \right) + \dots \right\} \label{2.17a} \\
~&~&\cr
 &~&\log (-z + i\varepsilon) ~=~ \log \vert z \vert + i\pi \, \theta(z)
\label{2.17b}
\end{eqnarray}
\end{subequations}
\begin{subequations}\label{2.18}
\begin{eqnarray}
S_{ij}(x) &=& \delta_{ij} \, S_j(x) \, K_j \label{2.18a} \\
~&~&\cr
S_j(x) &=&  {i\over {4\pi^2}} ~\left\{ {{2\gamma x}\over{(x^2-
i\varepsilon)^2}} + {m_j\over2} ~ {1\over{x^2-i\varepsilon}} \right. \cr
~&~&\cr
&-&~ m_j^2 \left.~\left[ ~{{\gamma x}\over{x^2-i\varepsilon}} ~+~
{\textstyle {1\over 4}} ~\log \left( - {\textstyle {1\over 4}} ~m_j^2\, x^2\,
\widetilde\gamma^2 + i\varepsilon \right) \right] + \dots \right\}
\quad . \label{2.18b}
\end{eqnarray}
\end{subequations}
Here $\widetilde\gamma = 1.781 \dots~$~.  Inserting the expansions
(\ref{2.17}), (\ref{2.18b}) in (\ref{2.16}) we obtain, putting $y=0$ in view
of translational invariance
\begin{eqnarray}
\Sigma ~&\cong& ~\hbox{\O} ~ \int d^4x \, \bar\psi(x) \, \gamma_5  \sum_j  K_j
~ \biggl\{ {{2\gamma x}\over{(x^2- i\varepsilon)^3}} ~+~ {m_j\over2} ~
{1\over{(x^2- i\varepsilon)^2}} \biggr. \cr
~&~&\cr
~&-& ~m_j^2 ~ {{\gamma x}\over{(x^2-i\varepsilon)^2}} ~-~ {\mu^2\over4} ~
{{\gamma x}\over{(x^2-i\varepsilon)^2}} ~\log \left(-~{ {1\over4}} ~ \mu^2 \,
x^2 \, \widetilde\gamma^2 + i\varepsilon \right) \cr
~&~&\cr
\biggl. &-& ~{{m_j^2} \over 4} ~ {1\over{x^2- i\varepsilon}} ~\log \left(-
~{ {1\over4}} ~ m_j^2 \, x^2 \, \widetilde\gamma^2 + i\varepsilon \right) +
\dots \biggr\} \gamma_5 \, \psi(0) \quad . \label{2.19}
\end{eqnarray}

From (\ref{2.13}) and (\ref{2.14}) we see that with the choice of the
constants $\kappa_i$ of (\ref{2.10}) the first two terms of the
``individual" propagators $S_j(x)$ cancel in the summation of
(\ref{2.19}) and do not appear in $\Sigma $. However, in view of our present
task we shall look at one of the propagators, $S_j(x)$, leaving the
cancellation to the end.

Following Ref.~\cite{1} we define $\bar\chi_\mu(x)$ such that there holds
\begin{subequations}\label{2.20}
\begin{eqnarray}
\bar\psi(x) &=& x_\mu \, \bar\chi_\mu(x) + \bar\psi (0) \label{2.20a} \\ ~&~&
\cr
&=& ~ x_\mu \, \bar\chi_\mu(x) + \int d^4x' \, \bar\psi(x') \,
\delta^4(x') \quad . \label{2.20b}
\end{eqnarray}
\end{subequations}
Consider now the second term of the left-hand side of (\ref{2.19}).
Inserting (\ref{2.20a}) in that term we have
\begin{eqnarray}
I_j ~&\cong& ~ {m_j\over 2} ~ \int d^4x \, \bar \chi_\mu(x) ~ {x_\mu \over {
(x^2 - i \eta)^2}} ~ \psi(0) \cr
~&~& \cr
&+& ~ {m_j\over 2} \int d^4x ~ \left( \int d^4x' ~ \bar\psi(x') \,
\delta^4(x') \psi(0) \right) ~ {1\over {(x^2-i \eta)^2}} \label{2.21}
\end{eqnarray}
where we have used the version (\ref{2.20b}) in order to make manifest the
contact character of this term.  The second term of (\ref{2.21}) shows that
$I$, and hence the individual terms of $\Sigma$, can contain contact
singularities with amplitude proportional to
\begin{equation}
\int d^4 x ~ {1\over {(x^2 - i\eta)^2}} \quad . \label{2.22}
\end{equation}
As shown in detail in Ref.~\cite{1}, Eq.~(\ref{2.22}) indeed is
logarithmically divergent. However, being a contact term it can contribute
only to the Wightman function, and not to the \hbox{S-matrix}.  At any rate,
here it is eliminated upon summation over $j$ in (\ref{2.19}), also for
the Wightman function. All remaining terms are regular for $x_\mu=0$ and
are best evaluated in momentum space.  Their sum yields convergent integrals,
individually for each $j$.

To summarize this Section, we have seen that in the  \hbox{g-functional}
formulation all matrix elements of the \hbox{S-matrix} are given by convergent
expressions and hence are mathematically rigorously defined, and the Green's
functions are those associated with the Lagrangian.  In this the present
formulation differs from the conventional regularization procedure where the
regularized Green's function is disconnected from the Lagrangian and where the
choice of the regularizing function is arbitrary and only dictated by
convenience.  At any rate, in view of K\"allen's theorem \cite{12} the
results of
the conventional procedure turn out to be ``correct" in the sense that they
agree with those obtained as rigorous solutions of the \hbox{g-functional}
formulation.
\section {QUANTUM ELECTRODYNAMICS}
\setcounter{equation}{0}
\renewcommand{\theequation}{3.\arabic{equation}}

As another example we now consider QED.  Thus we have to take care of charge
conservation, in addition to Lorentz invariance.  We take here the
\hbox{g-functional}
\begin{equation}
L \{ \psi \cdots \} ~=~ L_F(\lambda) ~+~L_B(\lambda) ~+~L_I(\lambda)
\label{3.1}
\end{equation}
with
\begin{eqnarray}
L_F(\lambda) &=& - \int~ d^4x ~:~ \left[
\bar \psi_1(x)
(\gamma\partial + m_1) \psi(x) ~-~ \bar \psi_2(x) (\gamma\partial
+ m_2) \psi_2(x) \right] ~:~ \label{3.2} \\
~&~&\cr
L_B &=& - \int~ d^4x ~:~ \left[ {1\over 4}~ F_{\mu\nu}(x)
F_{\mu\nu}(x) ~-~ \sum_\nu ~ \left( \partial_\mu B_\nu \partial_\mu
B_\nu ~-~ M^2 B_\nu B_\nu
\right) \right] ~:~ \label{3.3} \\
~&~&\cr
L_I &=& - i 4~ \int~d^4x ~:~ \left[ \left( e_1  \bar\psi_1 \gamma_\mu
\psi_1\right) ~+~ \left( e_2  \bar\psi_2 \gamma_\mu \psi_2 \right)
\right]~ \left( A_\mu + B_\mu \right) ~:~ \quad . \label{3.4}
\end{eqnarray}
and
\begin{equation}
F_{\mu\nu} ~=~ \partial_\mu \, A_\nu - \partial_\nu \, A_\mu  \quad .
\label{3.5new}
\end{equation}
This g-functional thus contains besides the Fermion and the photon field two
ghost fields: one Fermion and one Boson.  The latter is a mixture of spin~0
and spin~1.

We check for charge conservation.  To that end we introduce the
substitution
\begin{equation}
 \psi_1 ~\rightarrow ~ e^{i\alpha} ~\psi_1 \qquad , \qquad \psi_2
~\rightarrow ~ e^{i\beta} ~\psi_2 \quad , \label{3.5}
\end{equation}
with $\alpha,\beta$ infinitesimal, and we demand
\begin{equation}
{ {\delta L}\over{\delta \alpha}} ~=~ 0 \qquad , \qquad { {\delta
L}\over{\delta \beta}} ~=~ 0 \quad . \label{3.6}
\end{equation}
From Noether's theorem we have
\begin{equation}
\partial_\mu ~j_\mu^{(1)} ~=~0 \qquad , \qquad \partial_\mu ~j_\mu^{(2)} ~=~0
\quad , \label{3.7}
\end{equation}
\begin{equation}
j_\mu^{(n)} ~=~ i~ e_n ~\bar\psi_n ~\gamma_\mu~ \psi_n \quad , \label{3.8}
\end{equation}
i.e., separately conserved currents.  With the ansatz Eq.~(\ref{3.4}) here, in
contrast to the Pauli-Villars procedure, no photo-excitation from $\psi_1$ to
$\psi_2$ exists.  Therefore one is free to chose the ghost coupling constant,
$e_2$ at will.  Thus in the Fermion mass-renormalization graph
asymptotically at large loop momenta the Fermion propagator retains the
character $\mid p\mid^{-1}$.

The only place where the Fermion ghost contributes is in the vacuum
polarization, in that for every closed Fermion particle loop there exists an
analogous ghost loop.  The lowest order contribution then is
\begin{equation}
\Pi ~=~ \hbox{\O} \int \left[ {{e_1^2} \over {(p \!\!\!/-m_1)(q \!\!\!/-m_1)}}
~-
~ {{e_2^2}\over {(p \!\!\!/ -m_2)(q \!\!\!/ -m_2)}} \right] ~d^4p  \label{3.9}
\end{equation}
where
\begin{equation}
q ~=~p-k \label{3.10}
\end{equation}
and where the minus sign of the ghost contribution results from the ghost
field having the ``wrong" statistics, i.e., $\epsilon = -1$ in (\ref{2.5}).
In each of these terms the quadratic divergence is a contact singularity
and does not contribute to the open integral, while the surviving terms are
\begin{equation}
e_1^2 ~ \hbox{\O} \int ~d^4p ~{{p^2q^2 (m_1^2 - \zeta^2 m_2^2 ) ~+~ \cdots}
\over {(p^2 - m_1^2) (q^2 - m_1^2) (p^2 - m_2^2 ) (q^2 - m_2^2)}} \quad .
\label{3.11}
\end{equation}
Choosing
\begin{equation}
\zeta^2 ~=~ {{e_2^2}\over {e_1^2}} ~=~ {m_1^2\over m_2^2} \label{3.12}
\end{equation}
the integral is absolutely convergent.  With this choice in the limit
$m_2^2~\rightarrow ~\infty$ the Fermion ghost not only freezes out but also
decouples.

We now turn to the Boson sector.  The photon propagator is the usual one
(Feynman gauge)
\begin{equation}
\Delta_{ph} ~=~ \int ~ {{\delta_{\mu\nu}} \over {p^2 -i\eta}}~ d^4p
\label{3.13}
\end{equation}
while for the ghost it is directly the Klein-Gordon propagator
\begin{equation}
\Delta_G ~=~ \int ~ {{\delta_{\mu\nu}} \over {p^2 + M^2 - i\eta}}~ d^4p
\quad . \label{3.14}
\end{equation}
The full (photon + ghost) Boson propagator thus is
\begin{equation}
\Delta ~=~ -~\int {{\delta_{\mu\nu}~M^2} \over {p^2(p^2 - M^2) ~-~ i\eta}}~
d^4p \quad ; \label{3.15}
\end{equation}
this way together with the Fermion propagator $(p \!\!\!/-m)^{-1}$, all
graphs are absolutely
convergent for $|p| \to \infty$, and the limits $m_2^2,~M^2 \to \infty$
can be safely carried out.

\section {FORM OF THE STATIONARY STATE SOLUTIONS}
\setcounter{equation}{0}
\renewcommand{\theequation}{4.\arabic{equation}}

Having seen that the S-matrix can be computed in a mathematically solid
manner,
we now shall show that the solutions themselves are well defined and can be
used to construct the wave functions of the system, which then can be used to
evaluate its characteristics exactly as in the familiar non-relativistic
quantum mechanics.  It is here that the modification~(i) mentioned in the
Introduction becomes explicitly evident.

The computation of the \hbox{S-matrix} in the interaction picture is
based on a
time-dependent treatment.  Of course, this treatment can be applied to the
computation of stationary states \cite{19,20,21,22}.  On the other hand
a stationary
state is easiest to visualize in the time-independent treatment.  Therefore,
we
will begin by illustrating the form of the solution in the Schr\"odinger
picture using as examples the case of a single (physical) particle of the
Yukawa theory given above,
and that of a composite bound state system for the case of different
Fermions, for example the bound state of a muon and an electron.
We then will re-state the problem in the
interaction picture and derive a covariant eigenvalue equation in terms of
the
\hbox{U-matrix} elements.  Of course, for a convergent (or, within its range of
convergence, for a \hbox{semi-convergent}) theory the results are independent
of the choice of the method.  In particular, it is well known that the
Schr\"odinger-picture treatment is fully equivalent to the treatment by the
Feynman graph expansion, only being enormously more cumbersome for performing
an actual calculation.

In the Schr\"odinger picture the state vector of a given system, $\vert
W^{(S)}(t)\rangle$, obeys the equation
\begin{equation}
{\partial\over {\partial t}} ~ \vert W^{(S)}(t) \rangle ~=~ -i~ H (\lambda) ~
\vert W^{(S)} (t) \rangle \label{4.1}
\end{equation}
where
\begin{equation}
H (\lambda )~=~ H_0(\lambda) + H_1(\lambda) \label{4.2}
\end{equation}
is the Hamiltonian of the system; $H_0(\lambda)$ and $H_1(\lambda)$ are
associated with $L_0(\lambda) ~=~ L_F(\lambda) + L_B(\lambda)$ and
$L_I(\lambda)$, of Eqs~(\ref{2.1}) through (\ref{2.4}), respectively.  The
basic fields are those of the Heisenberg picture taken at the time $t=0$.
They
are taken to be solutions of $H_0( \lambda )$.  So, for example, the Fourier
decomposition of the Schr\"odinger-picture fields $\psi_i^{(S)}(x)$ is
\begin{eqnarray}
\psi_i^{(S)}(x) &=& \psi_i^{(+)(S)}(x) + \psi_i^{(-)(S)}(x) \cr
~&~&~ \cr
&=&  \int {{d^3p} \over {(2\pi)^{3/2}}} ~{\sqrt{ {m_i}\over
{\omega_i}}} ~ \left( b_{\bf p} \, u_{\bf p} \, e^{i{\bf px}} ~+~ d_{\bf
p}^\dagger \, v_{\bf p} \, e^{-i{\bf px}} \right) \label{4.3}
\end{eqnarray}
where we have suppressed the spin indices, and similarly for the other basic
fields of the Lagrangian.  Also, we use the notation $\omega_i ~=~+ {\sqrt
{{\bf p}^2 + m_i^2}}$.  In (\ref{4.3}) we have introduced the usual notation
for the positive frequency ($\psi^{(+)}$, annihilation operator) and negative
frequency ($\psi^{(-)}$, creation operator) fields.  In principle the basis
(\ref{4.3}) must be discretized in some manner, e.g., by Weyl's
eigendifferential method.  From now on we drop the superscript $(S)$
indicating the Schr\"odinger picture, and the explicit reference to the
parameter $\lambda$.

We note the commutation relations
\begin{eqnarray}
\left[ b_{\alpha {\bf p}}, \, b_{\beta{\bf q}}^\dagger \right]_{\epsilon} &=&
\xi\delta_{\alpha\beta} ~\delta^3 ({\bf p} -{\bf q}) \label{4.4} \\
~&~&\cr
\left[ d_{\alpha {\bf p}}, \, d_{\beta{\bf q}}^\dagger
\right]_{\epsilon} &=& \xi\delta_{\alpha\beta} ~\delta^3 ({\bf p} -{\bf q})
\label{4.5} \\
~&~&\cr
\left[ a_{\alpha {\bf p}}, \, a_{\beta{\bf q}}^\dagger
\right]_- &=& \xi\delta_{\alpha\beta} ~\delta^3 ({\bf p} -{\bf q})
\label{4.6}
\end{eqnarray}
where as above $ \xi~=~+1$  for particles and $=~-1$ for ghosts, and
$\epsilon~=~-1$ for the ``wrong" statistics.

A solution for a stationary state of a given system can be obtained by
expansion in the Fock-space representation, i.e., by constructing a complete
basis of states having the quantum numbers of the desired state (momentum,
spin, etc.) and diagonalizing the total Hamiltonian, Eq.~(\ref{4.2}), in that
basis.  The Schr\"odinger-picture state vector for the system consisting
of a single Fermion of momentum {\bf p} at infinite distance of any other
particles then will have the general form
(suppressing the spin indices)
\begin{equation}
\vert W(t) \rangle ~=~e^{-iEt} ~\vert W\rangle \quad  , \label{4.7}
\end{equation}
\begin{eqnarray}
 \vert W\rangle ~=~ C_{100} ~b_{\bf p}^\dagger     ~\vert
0\rangle ~&+& \int d^3k~C_{101}({\bf k})~b_{{\bf p}-{\bf k}}^\dagger
~a_{\bf k}^\dagger ~\vert 0\rangle \cr
~&~&~ \cr
&+& ~\int d^3k_1 \, d^3k_2 ~C_{102}({\bf k}_1{\bf k}_2)~
b_{{\bf p}-{\bf k}_1}^\dagger   ~a_{{\bf k}_1-{\bf k}_2}^\dagger ~a_{{\bf
k}_2}^\dagger ~\vert 0\rangle  + \ldots \cr
~&~&~ \cr
&+& ~\int d^3p_1 \, d^3p_2 ~C_{210}({\bf p}_1{\bf p}_2)~ b_{{\bf p}-{\bf
p}_1}^\dagger   ~b_{{\bf p}_1-{\bf p}_2}^\dagger ~d_{{\bf p}_2}^\dagger ~\vert
0\rangle  + \ldots \cr
~&~&~ \cr
&+& ~\ldots  \label{4.8}
\end{eqnarray}
where actually, e.g., $b_{\bf k}^\dagger = \sum_i b_{i{\bf k}}^\dagger$
to account for the auxiliary fields  as implied by (\ref{2.2}).
In (\ref{4.7}), $E$ is the energy, i.e., $E=+{\sqrt{{\bf p}^2+M^2}}$,
where $M$ is the at this juncture unknown mass of the system.

For the composite system we assume two distinct Fermions but only one Boson
species, and their assorted auxiliary particles.
The state vector for the composite system is similarly to above
\begin{eqnarray}
\vert W^{(c)}\rangle
&=& \int d^3p_1 C_{11000}({\bf p}_1) ~b_{{\bf p}-{\bf p}_1}^{(a)\dagger}
b_{{\bf p}_1}^{(b)\dagger}    ~\vert 0\rangle  \cr
&& \cr
&~&~+~ \int d^3p_1 \, d^3k ~C_{11001}({\bf p}_1{\bf k})~
b_{{\bf p}-{\bf p}_1}^{(a)\dagger}   ~b_{{\bf p}_1-{\bf k}}^{(b)\dagger}
~a_{\bf k}^\dagger ~\vert 0\rangle + \ldots \cr && \cr
&&~+~ \int d^3p_1 \, d^3p_2  d^3p_3
~C_{22110}({\bf p}_1{\bf p}_2{\bf p}_3)~ b_{{\bf p}-{\bf p}_1}^{(a)\dagger}
~b_{{\bf p}_1-{\bf p}_2}^{(b)\dagger}~b_{{\bf p}_2-{\bf p}_3}^{(a)\dagger}
~d_{{\bf p}_3}^{(a)\dagger} ~\vert 0\rangle  + \ldots \cr
&& \cr
&&~+~ \ldots \label{3.8a}
\end{eqnarray}

In the usual manner we now compute the Hamiltonian matrix by inserting
(\ref{4.8}) in (\ref{4.1}) and by multiplying on the left with the Hermitian
conjugate of one of the components   of the state vector at a time, for
example, $\langle \{\beta\} \vert = \langle 0\vert~b_{\bf p-k}\, a_{\bf k}$.
This way we  obtain the matrix equation
\begin{equation}
(E-E_{\{\beta\}})~C_{\{\beta\}} ~=~ \sum_{\{\alpha\}} ~\int d^3k_1~ \ldots
~C_{\{\alpha\}} \langle \{\beta\} \vert ~H_1~ \vert \{ \alpha\} \rangle
\label{4.9}
\end{equation}
where the fact has been used that $H_0$ is diagonal in the Fock space
configurations, yielding the value $E_{\{\beta\}}$ for the configuration
$\{\beta \}$.

The matrix (\ref{4.9}) now can be diagonalized, either directly,
i.e., \hbox{non-perturbatively,} or perturbatively, say, by the
Rayleigh -- Schr\"odinger method, as a power series in the coupling
constant, say $g$, i.e., by writing
\begin{equation}
E~=~E_0 + gE_1 + g^2E_2 + g^3E_3 + \ldots \quad  , \label{4.10}
\end{equation}
where for the case of semi-convergence the series has to be terminated.
(Equivalently, in the non-perturbative treatment the Hamiltonian matrix has to
be truncated.)   One has, for example,
\begin{equation}
g^2 \, E_2 ~=~ \sum{ \{\alpha \}} \int d\{\alpha \} ~ {{ \langle \{ 1 \}  \vert
~H_1~ \vert \{ \alpha \}\rangle ~ \langle \{ \alpha \}
\vert ~H_1~ \vert \{ 1 \}\rangle } \over {E_\alpha - E}} \label{4.11}
\end{equation}
where $ \{1\}$ is the ground configuration, i.e., the first term of
(\ref{4.8}).  The integration in (\ref{4.11}) is over the quantum numbers of
the configuration $\{ \alpha \}$, for example, the momentum ${\bf k}$ for the
configuration $\{\alpha\} ~=~ b_{{\bf p}-{\bf k}}^\dagger ~a_{\bf k}^\dagger $:
\begin{equation}
\int d \{ \alpha \} ~ {{ \langle \{ b_{\bf p}^\dagger  \} \vert ~H_1~ \vert \{
\alpha \}\rangle ~ \langle \{ \alpha \} \vert ~H_1~ \vert \{ b_{\bf
p}^\dagger \}\rangle } \over {E_\alpha - E}} ~=~  \int d^3k ~ {{ \langle \{
b_{\bf p}^\dagger \} \vert \,H_1 \, \vert b_{{\bf p}-{\bf k}}^\dagger \, a_{\bf
k}^\dagger \rangle \,\langle b_{{\bf p}-{\bf k}}^\dagger \, a_{\bf k}^\dagger
\vert H_1 \vert \{ b_{\bf p}^\dagger \} \rangle} \over {\omega_{\bf k} +
\omega_{{\bf p} - {\bf k}} - E}}  \quad . \label{4.12}
\end{equation}

In the conventional formulation integrals of this type diverge. Owing to
the auxiliary particles they are non-divergent in the \hbox{g-functional}
formulation;  thus Fock space is well-defined.   As long as the
configurations $\{ \alpha \}$, $\{ \beta \}$,
retained in the Hamiltonian matrix belong to those appearing as intermediate
states in the Feynman graphs connected with the terms of the convergent
part of the presumably \hbox{semi-convergent} series expansion for the
\hbox{S-matrix}, they
have physical significance.  If they belong to that part of the series
expansion where the terms have started to grow, their retention would tend to
diminish the accuracy of the results; and thus they should be omitted.  This
condition thus represents the criterion for the truncation of the secular
equation, Eq.~(\ref{4.9}), needed for the direct diagonalization.  (Further
details of the solution of field theory problems by diagonalization in the
Schr\"odinger picture, in particular the treatment of the relativistic
\hbox{center-of-mass}, have been presented in Refs. \cite{DG,DGQ}.)

Since the Feynman integrals converge and themselves are independent of $g^2$,
$E_2$ is independent of $g^2$, and $E_n$ is independent of $g^n$. Hence, the
eigenvalue $E$ of the solution joins smoothly to the non-interacting
theory.  More precisely:  if in fact the \hbox{S-matrix} expansion is only
semi-convergent, then as $g^2$ becomes smaller the beginning of the growth of
the terms in the expansion gets postponed, and for $g^2\to 0$ this beginning
occurs at arbitrarily large order $n$.  Thus, as mentioned in point~(j) in the
Introduction,  $E$, Eq.~(\ref{4.10}), and also the \hbox{S-matrix} itself, in
the \hbox{g-functional} formulation is analytic at the point $g^2=0$.

These analytic properties of the point $g^2~=~0$
are seen by the following limiting procedure:
Given a value of $\lambda$, and a value of $N$, then in
\begin{equation}
E^{N,\lambda}(g)~=~\sum^N~g^n~E_n^{\lambda}  \label{L1}
\end{equation}
there exists a $g_0~\ne~0$ such that for $|g|~\le~|g_0|$ and $n~\le~N$
there holds
\begin{equation}
n!~g^n~E_n^{\lambda}~\le~(n-k)!~g^{n-k}~E_{n-k}^{\lambda}~
 \label{L2}
\end{equation}
for positive integer $k$. Thus for $g~\to~0$ the value of
$E^{N,\lambda}(g)$ and its $n$-th derivatives with $n~\le~N$ are
independent of $N$.

We note the form of the wave function of the system.  It is given by
\hbox{[re-introducing} the superscript $(S)$]
\begin{equation}
\eta (x_i ,t) ~=~ \langle 0\vert~ \Psi^{(S)}~ \vert W^{(S)}(t) \rangle
\label{4.13}
\end{equation}
where for the single Fermion case $\Psi^{(S)}$ is the field
\begin{equation}
\Psi^{(S)} ~=~ \sum_{ \{\kappa\} }~ \prod_i^{i_{\scriptstyle \kappa}} ~
\bar\psi^{(+)}({\bf x}_i) ~\prod_j^{j_{\scriptstyle\kappa}} ~ \psi^{(+)}
({\bf
x}_j) ~ \prod_k^{k_{\scriptstyle \kappa}} ~ \varphi^{(+)} ({\bf x}_k) \quad .
\label{4.14}
\end{equation}
For the composite system the field is
\begin{equation}
\Psi^{(S)} ~=~ \sum_{ \{\kappa\} }~ \prod_i^{i_{\scriptstyle
\kappa}} ~\bar\psi^{(a)(+)}({\bf x}_i)
~\prod_j^{j_{\scriptstyle\kappa}} ~ \psi^{(a)(+)} ({\bf x}_j) ~
~ \prod_k^{i_{\scriptstyle \kappa}} ~ \bar\psi^{(b)(+)}({\bf x}_k)
~\prod_l^{l_{\scriptstyle\kappa}} ~ \psi^{(b)(+)} ({\bf x}_l) ~
~\prod_m^{m_{\scriptstyle \kappa}} ~ \varphi^{(+)} ({\bf x}_m)
\quad .
\label{3.14a}
\end{equation}

For the one-Fermion system only the following terms of the full set of
configurations, $ \{ \kappa \}$, are needed
\begin{eqnarray}
\Psi^{(S)}_{(1)} &=& \psi^{(+)} ({\bf x}_1)
~\left[ 1 + \varphi^{(+)} ({\bf y}_1) + \varphi^{(+)} ({\bf y}_1)
\varphi^{(+)} ({\bf y}_2) + \ldots \right] \cr
~&~& \cr
&~&~+~ \bar\psi^{(+)} ({\bf x}_1) ~ \psi^{(+)} ({\bf x}_2) ~ \psi^{(+)}
({\bf x}_3) ~\left[ 1 + \varphi^{(+)} ({\bf y}_1) + \varphi^{(+)} ({\bf y}_1)~
\varphi^{(+)} ({\bf y}_2) + \ldots \right] \cr
~&~&~ \cr
~&~&~+ ~  \ldots \quad . \label{4.15}
\end{eqnarray}
Analogously only a subset of the configurations of Eq.~(\ref{3.14a}) is needed
for the composite system.

According to (\ref{4.7}) and (\ref{4.8}) the wave function $\eta(x_i,t)$ of
(\ref{4.13}) has the overall time dependence $e^{-iEt}$ and is a mixture of
configurations containing different numbers of quanta of  the basic
fields.  This is fully analogous to the configuration mixing in atomic or
nuclear physics.  One may call $\Psi^{(S)}$ of (\ref{4.14}) the general
Schr\"odinger-picture solution field, and $\vert W^{(S)} (t)\rangle$,
(\ref{4.8}), the Schr\"odinger-picture solution state vector for the
\hbox{one-Fermion} system.  And, given that the basic
fields obey local commutation relations it is obvious that this is not
the case for the solution fields (\ref{4.14}) or (\ref{4.15}).  As can
easily be verified, the commutation relation $\left[ \Psi^{(S)},
\bar\Psi^{(S)} \right]_+$  -- which is the equal-time relation of the
corresponding interaction picture fields -- is a mixture of
\hbox{c-numbers}, creation operators, annihilation operators, and number
operators, and is non-local;
i.e., these fields do not commute at space-like separation, which is
contrary to one of the axioms of constructive field theory.  This way one
of the basic assumptions of Haag's theorem,\cite{4,5,6} i.e., that the
equal-time commutation relations of the fields $\Psi$ yield a local
\hbox{c-number}, is not fulfilled; hence, that theorem does not apply in
our \hbox{g-functional} formulation of QFT, and hence the interaction
picture exists for \hbox{non-trivial} theories.

As the last point we now indicate that the wavefunction can be computed
also in
a covariant manner.  To achieve this we now have to re-express (\ref{4.1}) in
terms of the \hbox{U-matrix}.  This will also show the connections with the
results of the previous Sections.  To that end we re-write our expressions in
the interaction picture.  We define for the state vector of the system in the
stationary state $n$ of energy $E_n$
\begin{equation}
\vert W_n^{(I)}(t) \rangle ~=~ e^{-iE_nt}~ |W_n^{(I)}\rangle ~=~ \sum_{ \{
\alpha\} } ~ w_{ \{ \alpha\} }^{(n)(I)}(t) ~ B_{\{\alpha\}}^\dagger
\vert0\rangle \quad . \label{4.16}
\end{equation}
Here we have compactified the notation of (\ref{4.8}) by introducing the
combined index ${ \{ \alpha\} } $.  The sum over ${ \{ \alpha\} } $ in
(\ref{4.16}) thus contains both the summation over the configurations and the
integration
(actually summation for a Weyl representation) over the momenta.  The product
of the creation operators of (\ref{4.8}) is denoted by $B_{\{ \alpha\}
}^\dagger$.  Similarly, we re-write eq~(\ref{4.14}) for the interaction-picture
fields as
\begin{equation}
\Psi^{(I)}(t) ~=~ \sum_{ \{\alpha\} }~ \prod_{abc}^{ \{\alpha \} }~
\bar\psi_a^{(+)}(t) ~ \psi_b^{(+)}(t) ~ \varphi_c^{(+)} (t) ~\equiv ~ \sum_{
\{\alpha\} } \, \Psi_{\{\alpha\}}^{(+)(I)}(x_i,t) \label{4.17}
\end{equation}
taken at the same time $t$ for all fields.  In (\ref{4.17}) the indices,
$a,b,c$, denote also the momenta in the field expansions according to
(\ref{4.3}), for example.

Herewith, we have
\begin{equation}
\eta_n(x_i,t) ~=~ \langle 0\vert ~\Psi^{(+)(I)} ~\vert W_n^{(I)}\rangle
\label{4.18}
\end{equation}
Being in the interaction picture the time-dependence of the products of
fields is
\begin{subequations}\label{19}
\begin{eqnarray}
\Psi_{\{ \alpha \} }^{(+)(I)}(x,t) &=& e^{- i\,E_{\{\alpha\}}t} \,
\Psi_{
\{ \alpha \} }^{(+)(I)}(x_i) \label{4.19a}\\
~&~& \cr
E_{\{\alpha\}} &=& \sum_i^{\{\alpha\}} ~ E_i \label{4.19b}
\end{eqnarray}
\end{subequations}
where the $E_i$ are the energies of the individual fields participating in
the configuration $\{\alpha\}$, eq~(\ref{4.17}).  Hence, the amplitudes of
the state vector for our state must have the time-dependence
\begin{equation}
w_{\{\alpha\}}^{(n)(I)}(t) ~=~ e^{-i(E_n-E_{\{\alpha\}}) t} ~
w_{\{\alpha\}}^{(n)(I)} \label{4.20}
\end{equation}
in order to achieve the overall time dependence of (\ref{4.16}).

We now follow the development of the state beginning at time $t_0$ by
means of the evolution operator U.  Dropping the superscript $(I)$ we have
\begin{equation}
| W_n(t_1)\rangle ~=~ U(t_1,t_0) ~ | W_n(t_0)\rangle \label{4.21}
\end{equation}
or, in full detail
\begin{equation}
w_{\{\alpha\}}^{(n)}(t_1) ~=~ \sum_\beta ~ U_{\alpha\beta}(t_1,t_0)
w_{\{\beta\}}^{(n)}(t_0) \quad . \label{4.22}
\end{equation}
Combining (\ref{4.22}) with (\ref{4.20}) we obtain
\begin{equation}
e^{-i(E_n-E_{\{\alpha\}})(t_1-t_0)}~ w_{\{\alpha\}}^{(n)} ~=~ \sum_\beta
U_{\alpha\beta}(t_1,t_0) \, w_{\{\beta\}}^{(n)} \quad , \label{4.23}
\end{equation}
dropping the argument $t_0$ in the amplitudes. We now introduce the operator
\begin{equation}
e^{-i \int_{t_0}^{t_1} (E_n-H_0)dt'} ~=~ U^{(0)}(E_n;t_1,t_0) \label{4.24}
\end{equation}
which is diagonal in our basis.  It has the structure of a mass counter term.
Herewith, (\ref{4.22}) can be written as
\begin{equation}
\sum_n ~ \left[ \delta_{\alpha\beta} \, U_{\alpha\beta}^{(0)}(E_n;t_1-t_0) -
U_{\alpha\beta} (t_1-t_0)  \right] ~ w_{\{\alpha\}}^{(n)} ~=~0
\label{4.25}
\end{equation}
which is a (non-linear) eigenvalue equation for the energy and the state
vector amplitudes.  Upon the limit $t_0\to -\infty$, $t_1\to +\infty$, one
arrives at the covariant expressions for the \hbox{S-matrix} elements which
can be computed by the familiar Feynman graph expansion.  This way we see
that the concept of the wavefunction of a system is not limited to
non-relativistic physics but has a precise meaning also in the frame of
relativistic \hbox{g-functional} QFT.

\section {SUMMARY}
\setcounter{equation}{0}
\renewcommand{\theequation}{5.\arabic{equation}}

In this paper we have demonstrated that in the \hbox{g-functional} formulation
of quantum field theory every step leading from the Lagrangian to
the \hbox{S-matrix} can be defined unambiguously in a mathematically rigorous
way; the renormalizations are non-divergent and can be carried through
explicitly; and the state vectors are normalizable.  As mentioned in point (E)
of the Introduction, concerning the convergence properties of QFT we assume
\hbox{semi-convergence} of the perturbation expansion. The solution fields are
non-local and hence Haag's theorem does not apply; i.e., the interaction
picture exists.  The state vectors of the solutions are linear superpositions
of Fock space configurations.
This is true both for single particles, e.g., electrons, and for composite
systems, e.g., positronium.
For a small coupling constant, in the solution the ground configuration has
the largest amplitude; the higher configurations are
corrections.  The solutions thus connect smoothly with the ``free"
\hbox{theory}; i.e., for a fixed $\lambda$ the limit $g^2 \to 0$ within the
limitations imposed by the possible semi-convergence of the expansion is
analytic in the usual sense: for given $\epsilon$ and $n$, there is a $g_0^2$
such that for $g^2 < g_0^2$ the $n^{th}$ configuration has an admixture
$ \vert A_n \vert^2 < \epsilon$.  Herewith we have shown that non-trivial
relativistic QFT is not only consistent mathematically, but also makes sense
from the point of view of physics.  Thus, the theory is ``transparent," i.e.,
from contemplating the solutions one can draw conclusions concerning the
underlying Lagrangian.  This connection is not colored by the mathematics:
the result does not depend on the choice of the \hbox{g-functional}, as
long as it has the desired Lagrangian as the limiting form, and it has the
needed convergence properties.  Conversely, one then is on safe grounds to
judge whether a chosen Lagrangian, i.e., always defined as the
limit of a \hbox{g-functional}, is suitable for describing the physics at hand.
Thus we can conclude that non-trivial relativistic quantum field theory exists
as a rigorous mathematical theory and that it can be employed in attempting
the description of Nature.

\setcounter{equation}{0}
\renewcommand{\theequation}{A.\arabic{equation}}

\begin{center}{\bf  APPENDIX A}
\end{center}

\bigskip

\noindent
{\bf \hbox{g-functions} and \hbox{meta-g-functions}}

Generalized functions (\hbox{g-functions}), also called distributions, are
defined in a limit procedure by their action on a function, the ``test
function."  In the terminology of Lighthill \cite{23} the limit procedure is
carried out in terms of sets of ``good functions" which contain a parameter,
$\lambda$,  and the \hbox{g-functions} arise in the limit $\lambda\to 0$.
Considering as an example the  $\delta-$ function, we have
\begin{equation}
\delta(x) ~=~ \lim_{\lambda\to0} \Delta_\lambda(x) \label{A.1}
\end{equation}
and
\begin{equation}
\lim_{\lambda\to 0} \int \Delta_\lambda(x) \,f(x)\, dx ~=~f(0)
\quad . \label{A.2}
\end{equation}
Of course, (\ref{A.1}) is only a symbolic notation and to be understood in
the context of (\ref{A.2}).  There exist unlimitedly many different choices
for the sets  of ``good functions" $\Delta_\lambda(x)$  which can be used.
But all of them make mathematical sense only for $\lambda\neq 0$.  The
functions $\Delta_\lambda(x)$ are meaningless for $\lambda=0$.  The limit
procedure thus is to be carried over the domain $[\lambda_0,0)$ for
$\lambda$, i.e., the domain open at $\lambda=0$, and where $\lambda_0$ is
some arbitrary, ``small" value.  Also, given a set  $\Delta_\lambda(x)$, the
functions $f(x)$, the ``test functions," must have certain characteristics
for (\ref{A.2}) to hold.

All this is well-known; the mathematics of \hbox{g-functions} is essentially
complete \cite{23,24}.  The most essential of their characteristics is that
their action, analogous to (\ref{A.2}), is unambiguous, and that they have
unambiguous Fourier transformations.  Thus the Fourier transform of (\ref{A.2})
is given by the convolution of the Fourier transforms of $\Delta_\lambda(x)$
and of $f(x)$.

When attempting to apply (\ref{A.2}) to the case
\begin{equation}
f(x) ~=~ {1\over x}~ g(x) \simeq \left( {P\over x} ~+~ Z \,\delta(x) \right)
\,g(x)         \label{A.3}
\end{equation}
[$g(x)$ a ``good" function] one arrives formally at
\begin{equation}
\lim_{\lambda\to 0} \int \Delta_\lambda(x)~ {1\over x} ~g(x) \,dx
~=~\lim_{\lambda\to 0} \int \Delta_\lambda(x) \left( {P\over x}~ +~ Z\,
\delta(x) \right) \,g(x) \,dx  \label{A.4}
\end{equation}
which breaks the frame of \hbox{g-functions}; it involves the product of two
\hbox{g-functions}.  Such quantities have been
called ``\hbox{meta-g-functions}." \cite{1}  The  Wightman functions of QFT
frequently contain \hbox{meta-g-functions}.

Going back to (\ref{A.1}) one finds
\begin{equation}
[\delta(x)]^2 ~=~ \lim_{\scriptstyle\lambda_1\to 0 \atop
\scriptstyle\lambda_2\to 0} \Delta_{\lambda_1}(x) \, \Delta_{\lambda_2}(x)
\quad .   \label{A.5}
\end{equation}
One has here two independent limits.  The result one would obtain from
\begin{equation}
\lim_{\lambda_1,\lambda_2\to 0} \int \Delta_{\lambda_1}(x) \,
\Delta_{\lambda_2}(x) \, g(x) \, dx  \label{A.6}
\end{equation}
depends on the manner one performs the limit procedure.  Thus (\ref{A.6}) is
inherently ambiguous.  \hbox{Meta-g-functions} unavoidably yield ambiguous
results.  Not much is known about their properties.  Still, one can define
their ``singularity character," which is independent of the way one performs
the  limits, e.g., in (\ref{A.6}).  Thus one may call a \hbox{meta-g-function}
$D^{(n)}(x)$ to be of degree $n$ ($n=$integer) if there holds
\begin{equation}
\int ~D^{(n)}(x)~ x^{n+k} \, dx ~=~ 0 \label{A.7}
\end{equation}
with integer $k \geq 0$.  With (\ref{A.7}) one sees that the general solution
of the  equation (measure $dx$)
\begin{equation}
x^n ~f(x) ~=~ g(x) \label{A.8}
\end{equation}
in addition to \hbox{g-functions} ($\delta-$functions and their derivatives)
contains also \hbox{meta-g-functions} of degree up to $n$.  (The
Eq.~(\ref{A.8})
can be used to define  \hbox{meta-g-functions} of degree $n$ \cite{1}].
An example
of a degree 2 \hbox{meta-g-function} is Eq.~(\ref{A.6}).
\hbox{Meta-g-functions} of degree $n~=~1$ are \hbox{g-functions}.)  Some
further discussion on the properties of \hbox{meta-g-functions} is
contained in
Ref.~\cite{1}.

\bigskip

\noindent
{\bf Contact type g-functions and meta-g-functions}.

The open integral, $\hbox{\O} \int_a^b\,dx\,\cdot$, is defined to leave out
the single point $x=0$. Precisely, it is defined as a
definite integral over the range $\hbox{\O}(a,b)=(a,0),(0,b)$.  It differs
from the integral $\int_a^b\,dx\,\cdot$
which has the range $(a,b) = (a,0]\,(0,b) = (a,0)\,[0,b)$.  In order for these
two integrals to be different, the integrand must be a (\hbox{meta-)g-function}
of the contact type, for example, $\delta(x)$, or $[\delta(x)]^2$, etc.  Thus,
for example,
\begin{equation}
\hbox{\O} \int dx \, [f(x)\, \delta(x) + g(x)] = \int dx \, g(x) \label{A.9}
\end{equation}
for $f(x)$ and $g(x)$ (ordinary) test-functions.  The \hbox{g-function}
$\delta'(x)$ is not of the contact type as can be seen from the definition
\begin{equation}
\delta '(x) ~=~ \lim_{\eta\to 0} ~ {{\delta(x+\eta) - \delta(x-\eta)} \over
{2\eta}} \quad . \label{A.10}
\end{equation}
It is retained in the open integral.  Further details are given in
Ref. \cite{1}.

We could carry out our analysis of the \hbox{meta-g-functions} because we used
the procedure of Lighthill \cite{23}.  The Schwartz -- Gel'fand
definitions \cite{24}
do not lend themselves to the required generalizations.
\bigskip
\noindent
{\bf \hbox{g-functionals} and \hbox{meta-g-functionals}}

In analogy to the \hbox{g-functions} one can define generalized functionals,
for  short \hbox{g-functionals}.  They are functionals of the fields $\psi(x)$,
which can be \hbox{c-numbers} or operators.  In analogy to the good functions
generating the \hbox{g-functions}, the functionals generating the
\hbox{g-functionals}
contain a parameter $\lambda$, and they are well-defined for
$\lambda \neq 0$; that means that the functional equations they are associated
with (e.g., the equations of motion) have solutions for $\psi$ (and for the
state vectors upon which the field operators act) for $\lambda\neq 0$.  They
make no mathematical sense for $\lambda = 0$.  The \hbox{g-functionals} are
then defined in analogy with the \hbox{g-functions}: in order to achieve
mathematically sound results the calculation is performed for a non-zero, fixed
value of $\lambda$.  Only after the completion of the calculation one goes to
the limit $\lambda\to 0$, in the strict sense that $\lambda = 0$ is not
permitted; the limit procedure is to be carried out over the domain
$[\lambda_0,0)$, which is open at $\lambda = 0$, with arbitrary ``small"
$\lambda_0$.  To emphasize: the limit procedure is carried out for the
{\it solutions} of the theory (e.g., the \hbox{S-matrix}), i.e., the solutions
obtained for ``small"~$\lambda$.

Furthermore, in complete analogy to \hbox{g-function} theory one can give for
typical, general cases of \hbox{g-functionals} rules, for example, Feynman
rules, on how to write the results  of the $\lambda\to 0$ limit procedure.
And, in the same way as for \hbox{g-functions}, one always can check the rules
by actually computing the $\lambda\to 0$ results.

As is well-known, the solutions of the \hbox{g-functional} theories contain
\hbox{g-functions}.  For example, the Feynman integrals, say for QED, contain
light-cone $\delta-$functions and their derivatives.  These \hbox{g-functions}
survive the $\lambda\to 0$ limit procedure.
One may say that \hbox{g-functionals} lead to \hbox{g-functions}.
Inasmuch as \hbox{g-functions} have well-defined characteristics, the
solutions
of \hbox{g-functional} theories are also fully, unambiguously defined.

One also can introduce the concept ``meta-generalized-functionals" arising in
the context of \hbox{g-functionals} as analogues
of the  \hbox{meta-g-functions} which  arise in the
context of \hbox{g-functions}.  In the same way that \hbox{meta-g-functions}
are inherently ambiguous, the \hbox{meta-g-functionals} are inherently
ambiguous.   Examples for the \hbox{meta-g-functionals}
are the ``non-renormalizable" quantum field theories: they cannot be made
unambiguous -- even by a $\lambda\to 0$ limit procedure.  In fact, the Feynman
integrals arising in \hbox{non-renormalizable} theories
contain non-contact-type \hbox{meta-g-functions}, which belong in the
solutions \cite{1} and which render the solutions ambiguous.  Again,
one may say
that \hbox{meta-g-functional} theories lead to \hbox{meta-g-functions}.

\bigskip
\bigskip

\setcounter{equation}{0}
\renewcommand{\theequation}{B.\arabic{equation}}

\centerline
{\bf APPENDIX B}

\bigskip

\noindent
{\bf Negative Metric Fields}

Consider the action
\begin{equation}
L ~= \int \bar\psi(x) \,(\gamma\partial+m) \, \psi(x)~ d^4x \quad . \label{C.1}
\end{equation}
The Fourier expansion of the fields is
\begin{equation}
\psi(x) ~= \int {{d^3p}\over{(2\pi)^{3/2}}} ~ {\sqrt{m\over E_p}}~ \left[
b_{\bf p} \, u({\bf p}) ~ e^{i({\bf px}-Et)} ~+~ d_{\bf p}^\dagger \, v({\bf
p}) ~ e^{-i({\bf px}-Et)} \right] \quad . \label{C.2}
\end{equation}
The Fock operators obey the usual anti-commutation relation except for the
``wrong" sign:
\begin{equation}
\left[ b_{\bf p} ,b_{\bf p'}^\dagger \right]_+ ~=~ \left[ d_{\bf p}, d_{\bf
p'}^\dagger \right]_+ ~=~ -\delta^3({\bf p} - {\bf p}') \quad . \label{C.3}
\end{equation}
The Hamiltonian here is
\begin{equation}
H ~=~ -\int {{d^3{\bf p}}\over {(2\pi)^{3/2}}} ~ E_p ~ \left\{ :
b_{\bf p} ^\dagger b_{\bf p} : ~+~ : d_{\bf p}^\dagger  d_{\bf p}: \right\}
\quad . \label{C.4}
\end{equation}
The equations of motion are,
\begin{eqnarray}
[H,\psi(x)]_- &=& -\int {{d^3p}\over {(2\pi)^{3/2}}} \int {{d^3k}\over
{(2\pi)^{3/2}}}~ E_p \left\{ \left[ : b_{\bf p} ^\dagger b_{\bf p} :  b_{\bf k}
~-~ b_{\bf k} : b_{\bf p}^\dagger  b_{\bf p} : \right] \,
u_{\bf p} \, e^{i({\bf px}-E_pt)} \right. \qquad \qquad  \cr
~&~& \cr
&~& {\phantom{-\int {{d^3p}\over {(2\pi)^{3/2}}} \int {{d^3k}\over
{(2\pi)^{3/2}}} }} \left. + \left[ : d_{\bf p} ^\dagger d_{\bf p} :
d_{\bf k}^\dagger ~-~ d_{\bf k}^\dagger : d_{\bf p}^\dagger  d_{\bf p} :
\right] \, v_{\bf p} \, e^{-  i({\bf px}-E_pt)} \right\} ~~ .~~  \label{C.5}
\end{eqnarray}
We have
\begin{eqnarray}
: b_{\bf p}^\dagger b_{\bf p} :~ b_{\bf k}-b_{\bf k} ~
: b_{\bf p}^\dagger b_{\bf p} : &=& :b_{\bf p}^\dagger b_{\bf p}
:~b_{\bf k} - \left( -b_{\bf k} b_{\bf p}^\dagger - b_{\bf p}^\dagger b_{\bf k}
+ b_{\bf p}^\dagger b_{\bf k}\right) b_{\bf p} \cr
~&~& \cr
&=& :b_{\bf p}^\dagger b_{\bf p} :~b_{\bf k} - \delta^3( {\bf k}-{\bf p})
b_{\bf k} + b_{\bf p}^\dagger b_{\bf k} ~ b_{\bf p} \cr
~&~& \cr
&=& :b_{\bf p}^\dagger b_{\bf p} :~b_{\bf k} - \delta^3(k-p) b_{\bf k} -
:b_{\bf p}^\dagger b_{\bf p}: b_{\bf k} \cr
~&~&\cr
&=& - \delta^3( {\bf k} - {\bf p})~ b_{\bf k}  \label{C.6}
\end{eqnarray}
and similarly for the other terms.  Thus
\begin{eqnarray}
[H,\psi(x)]_- &=& -\int {{d^3p}\over{(2\pi)^{3/2}}} ~ E_p \left\{ b_{\bf p} \,
u_{\bf p} \, e^{ i({\bf px}-E_pt)} - d_{\bf p}^\dagger \, v_{\bf p} \, e^{-
i({\bf px}-E_pt)} \right\} \cr
~&~& \cr
&=& {1\over i} ~ {\partial\over\partial t} ~ \psi(x) \label{C.7}
\end{eqnarray}
which agrees with the equations of motion for positive metric fields.  The
expressions for the interaction picture development thus remain unchanged.

\newpage \noindent {\large\bf ACKNOWLEDGMENT}

I would like to thank Y.~Nambu and G.~Efimov for very useful discussions,
M.~Oberguggenberger for a very informative  correspondence concerning the
\hbox{g-functions} and \hbox{meta-g-functions}, and E.~Marx for a careful
reading of the manuscript and helpful comments.


\begin{thebibliography}{99}

\bibitem{4}  R. Haag, On Quantum Field Theory, Kgl. Dansk Mat.-Fys. Medd.
            {\bf 29}, 12   (1955).

\bibitem{5}  R. F. Streater and A. S. Wightman, {\it PCT, Spin \& Statistics
              and All That}, W.~A.~Benjamin Inc. New York, Amsterdam (1964).

\bibitem{6}  P. Roman, {\it Introduction to Quantum Field Theory}, John Wiley
               \& Sons, New York, London, Sydney, Toronto (1969).

\bibitem{7}  M. Danos, in {\it Interaction Studies in Nuclei}, H.~Jochim and
              B.~Ziegler, Eds., p.~885, North Holland Publ. Co., Amsterdam
          (1975).

\bibitem{8}    S. L. Adler, Phys. Rev. {\bf 177}, 2426 (1969).

\bibitem{9}    J. S. Bell and R. Jackiw, Nuovo Cimento {\bf 60A}, 47 (1969).

\bibitem{1}  M. Danos and J. Rafelski, Nuovo Cimento {\bf 49A}, 326 (1979).

\bibitem{2}  M. Danos and L. C. Biedenharn, Phys. Rev. {\bf D36}, 3069
(1987).

\bibitem{3}  See also T. D. Kieu, Phys. Rev. {\bf D44}, 2584 (1991).

\bibitem{10}  N. N. Bogoliubov, A. A. Logunov and I. T. Todorov, {\it
Introduction
               to Axiomatic Field Theory}, W. A. Benjamin Inc., Reading,
               Massachusetts (1975).

\bibitem{11}  N. N. Bogoliubov, A. A. Logunov, A. I. Oksak and I. T. Todorov,
               {\it Obshchie Printsipie Kvantovoi Teorii Polya}, Nauka, Moscow
         (1987)   (in Russian).

\bibitem{12}   G. K\"allen, Helv. Phys. Acta {\bf 25}, 417 (1952).

\bibitem{VAR}  M. Danos, Ward -- Takahashi Identities and Noether's Theorem
in Quantum Field Thearies, to be published; preprint hep-th 9702096

\bibitem{13}  T. D. Lee and G. C. Wick, Nucl. Phys. {\bf B9}, 209 (1969); {\bf
               B10}, 1 (1969);    Phys. Rev. {\bf D2}, 1033 (1970).

\bibitem{14}   D. L. Nordstrom, Phys. Rev. {\bf D4}, 1611 (1971).

\bibitem{15}  M. Danos, W. Greiner and J. Rafelski, Phys. Rev. {\bf D6}, 3476
                (1972).

\bibitem{16}  M. Danos, W. Greiner and J. Rafelski, Z. Physik {\bf 258}, 147
               (1973).

\bibitem{17}  J. Polchinski, Nucl. Phys. {\bf B231}, 269 (1984).

\bibitem{18}      G. Keller, C. Hopper and M.~Salmhofer, Perturbative
             Renormalization and effective Lagrangian in $\Phi_4^4$,
          Preprint MPI-PAE/PhT 65/90 (November 1990).

\bibitem{19}   M. Gell-Mann and F. Low, Phys. Rev. {\bf 84}, 350 (1951).

\bibitem{20}   E. E. Salpeter and H. A. Bethe, Phys. Rev. {\bf 84}, 1232
(1951).

\bibitem{21}   D. Lurie, {\it Particles and Fields}, Interscience Publishers,
New York, London, Sydney (1968).

\bibitem{22}   G. T. Bodwin, D. R. Yennie and M. A. Grogorio, Rev. Mod. Phys.
{\bf 57} Part~I,  723 (1985).

\bibitem{DG}  M. Danos and V. Gillet, Relativistic many-body bound systems,
NBS Monograph 149 (1975).

\bibitem{DGQ} M. Danos. V. Gillet and M. Cauvin, {\it Methods in Relativistic
Nuclear Physics}, North Holland, Amsterdam (1984).

\bibitem{23}   M. J. Lighthill, {\it Introduction to Fourier Analysis and
Generalised Functions}, Cambridge University Press, London (1958).

\bibitem{24}   I. M. Gel'fand and G. E. Shilov, {\it Generalized Functions},
Academic Press, New York, London (1964).


\end{thebibliography}
\end{document}